\documentclass[proof]{WileyASNA-v1}

\articletype{Article Type}%

\received{ }
\revised{ }
\accepted{ }

\usepackage{xcolor}

\newcommand{\refe}{}

\begin{document}
\UseRawInputEncoding
\title{QPOs are more than timing features: Applying the Accretion-Ejection Instability in Super massive black-holes}

\author[1,2]{Peggy Varniere*}



\address[1]{\orgdiv{Universit\'e  Paris Cit\'e, CNRS}, \orgname{Astroparticule et Cosmologie}, \orgaddress{\state{F-75013 Paris}, \country{France}}}

\address[2]{\orgdiv{Université Paris-Saclay, Université Paris Cité, CEA, CNRS}, \orgname{AIM}, \orgaddress{\state{F-91191 Gif-sur-Yvette}, \country{France}}}


\corres{\email{varniere@apc.univ-paris7.fr}}


\abstract{Low Frequency QPOs (LFQPOs) have always been seen uniquely as a timing feature, with most models focused on reproducing only their timing behavior. Previously, the Accretion-Ejection Instability (AEI) predicted the existence of a more subtle effect on their energy spectrum. In the case of LMXB, that effect was deemed impossible to detect.
While the AEI can be naturally expanded to higher-mass systems, there was no observational drive to do it until now. However the several (candidate) QPOs detected in supermassive black hole sources in the last decade, now seems the right moment to explore the applicability of the AEI model. Using our numerical observatory, we compute the different observables for a system with the AEI active and 
we take advantage of the longer timescale coming from the higher mass system, to see if the effect on the energy spectrum hinted at in LMXB, could become detectable.
After confirming that the AEI causes an imprint of the QPO on the energy spectrum that should reach detectable levels in SMBH, we turned to observations of 2XMM J123103.2 and performed a proof of concept spectro-timing fit of the QPO pulse profile simultaneously with how its energy spectrum changes along it.
\refe{While the limited 2XMM J123103.2 data showed the predicted effect at only $1.7\sigma$, it is fully consistent with the AEI model and hence worthy of further investigation in other systems.}
}
\keywords{black hole physics, accretion disks, magnetohydrodynamics (MHD)}

%

\maketitle



\section{Introduction}
%
 
   Since the beginning of X-ray observations we have seen how variable black-hole systems are, with, in particular, the fast quasi-periodic variations that are 
  known as Quasi-Periodic Oscillations  \citep[QPO, see for example][for a review]{Remillard06}. 
  In Low Mass X-ray Binaries (LMXB) with a black hole those QPOs are generally separated in two categories based on their behavior and 
  frequency, namely the High-Frequency QPO (HFQPO) between $40-450$ Hz and the Low-Frequency QPO (LFQPO)  in the range $0.01-30$ Hz. While HFQPOs 
  are rather rare and with a low rms amplitude of a few percent, LFQPOs are more common, to the point of being dubbed \lq ubiquitous\rq and can attain a high rms 
  amplitude  of up to $30$\%.
  Those LFQPOs are studied in the time domain where they are easily identifiable as a peak in the Power Density Spectrum (PDS) and are often shown to correlate with the 
  spectral evolution of the source\citep{rodriguez02}.

  \refe{
      \noindent   While we have a lot of observations in those low mass black-hole systems, no consensus on what causes the QPO modulation has been reached.
     All the proposed models can be loosely separated in different families such as the one based on the Lense-Thirring precession~\citep{stella98,schnittman06,ingram09,nixon14,Veledina15} or 
   different instabilities or wave propagations in either the disk or the corona \citep{TP99,chakrabarti00,titarchuk04,cabanac10,oneill11}.
 For a more in-depth presentation of those different models and the related observations we refer to the general review on X-Ray Properties of Black-Hole Binaries \citep{Remillard06} and their QPOs \citep{Ingram19}.
While all of those models have successfully compared with observations at some degree, those comparisons were always based on differents and limited subsets of the overall observed QPOs' characteristics, 
hence the lack of a consensus on the origin of the modulation. 
}

%
     Even though LFQPOs have a high rms amplitude, the presence 
     of what causes this strong modulation of the flux is ignored when analysing the simultaneous energy spectrum. Among the 
     reasons for this disregard is the fact that what we observe is not spatially resolved and averaged out over many of the QPO cycles which 
     makes it difficult to disentangle what could be the QPO from the other components of the system. 
     On top of that, it is not necessary to add another component to get a good enough fit of the spectral energy distribution. 

     While this has been the case since the beginning of Xray timing, it all changed when we started finding QPOs in Super Massive Black-Hole systems as the longer timescale of those QPO 
     can be used in our favor to lift the restriction of pure timing for the QPO and we can start exploring potential impacts on the energy spectrum. \\

     This is especially interesting within the framework of the Accretion-Ejection Instability whose non axi-symmetric structure has been shown to imprint the energy 
     spectrum\citep{VB05},  even though we only have hints of this effect through indirect detection in LMXB\citep{VMR16}. In the next section we will briefly present the 
     basics of the AEI and some of the previous comparisons with observations made in low mass systems. We will then turn to characterizing the AEI-born QPOs in super massive black-hole systems,
     focusing on aspects unreachable in faster systems. Lastly, we will present the case of  2XMM J123103.2 and use its QPO as a proof of concept of spectro-timing analysis 
     at the scale of the QPO modulation.

\section{the AEI as a model for LFQPOs}

      The model based  on the Accretion-Ejection Instability \citep[AEI,][]{TP99}  fits  in the category of  \lq instability-generated\rq\  QPO where the 
       \lq criteria\rq\ of the instability is the reason for the on and off behavior of the QPOs.      
       This instability was proposed to explain LFQPOs in LMXBs more than two decades ago and 
        \refe{since then several of its characteristics have been  successfully compared with observations from multiple sources.}


    \subsection{Basics of the AEI}
 
      The AEI occurs in the inner region of a fully magnetized accretion disk, namely when the magnetic field is close to equipartition with the gas pressure. While the AEI does not require the 
      central object to be a black hole, the strong gravity present in those objects impacts its behavior when the inner edge of the disk comes close to its last stable orbit 
      \citep{varniere02,VTR12}.
      The AEI is a global instability taking the form of a spiral wave in the inner region of the disk. In this model, the observed QPO modulation is linked with the existence and 
      amplitude of the spiral \citep{2005AIPC..797..631V,VV16}. 
       At  the location of the corotation between this spiral wave and the disk fluid,
      a Rossby vortex forms. When taking into account the low density corona surrounding the system, this vortex will twist the foot-point of the field lines and thus emit upward
      the energy and angular momentum stored in it. The name of the instability hence comes from the fact that accretion and ejection are linked through it \citep{VT02}.

    \subsection{Previous comparison with observations in low mass system}

         As with most of the LFQPO models, the AEI was first compared mostly to the frequency of the observed QPOs with the 
         specificity of trying to explain the two different signs of correlation observed between the fit-inferred inner disk position and the LFQPO frequency in different objects\citep{rodriguez02}.
         Indeed, taking into account relativistic effects on the AEI,  we found a reversal of correlation between the frequency of the 
         LFQPO and the position of the inner edge of the disk when it gets close to the last stable orbit of the black-hole \citep{varniere02}.
         This work was later expanded by finding observationally an object showing a reversal of correlation and comparing it with the prediction of the AEI \citep{mikles09}. \\
         

        While doing the theoretical development of the AEI as a model for the LFQPOs we also did more direct comparisons with observations, every time trying to pinpoint 
        observational consequences of unique characteristics of the AEI. 
        In order to compare more directly with observations, we created a fit-model based on the AEI
        which we then used to fit the QPOs of the 98-99 outburst of XTE J$1550$-$564$ \citep{VV17} hence demonstrating that the AEI was able to reproduce the evolution
        of the QPOs parameters (frequency, rms and Q) during an outburst. As of now, it is the only LFQPO model that has done such fit. 
\newline
      
        Even if the AEI was  proposed to exist in the inner region of an accretion disk around a low-mass black hole, the mass of the central object only impacts the scaling of the systems 
        not the instability itself. It is therefore natural to want to expand it to SMBHs and look for similar QPOs in those systems as a new way to test the model. 
        \refe{Indeed, as we go to higher masses and therefore longer timescales, it might become possible to look at the behavior of the source \textit{along} the phase of the QPO rather
        than integrated over the full oscillation.}

 \section{AEI-born QPO characteristics in SMBH}      
 \label{sec:AEI_SMBH}
 
 \refe{ While the previous section concentrated on 
 low mass systems, we are now focusing on computing the lightcurves and energy spectra  from higher mass systems with the AEI active 
 in order to see what could be learned from them.}

\subsection*{GR ray tracing of the impact of a spiral from the AEI}
\label{sec:toymodel}
           In order to explore more qualitatively the impact of going to higher mass we will use a similar approach as  in \citet{VV16} with a simple model, coming from fitting the
       numerical simulation of the AEI~\citep[see for example][]{VTR12}, representing the impact of the instability on the disk and then ray traced the emission back to the observer
       with GYOTO\footnote{{GYOTO can be downloaded at http://gyoto.obspm.fr}}~\citep{Vin11}.
       We consider a disk around a black-hole of mass $m$ and spin $a$ with the temparature profile being fitted from AEI simulations taking the form of:
   \begin{eqnarray}
   \label{eq:T}
 T(t,r,\varphi) &=& T_0(r) \\ \nonumber
 &&\times\left[  1 + \gamma \left(\frac{r_c}{r}\right)^\beta\mathrm{exp}\left(-\frac{1}{2}\left(\frac{r-r_s(t,\varphi)}{\delta \,r_c}\right)^2\right)\right]^2 \\ \nonumber
  \end{eqnarray}     
       Where the continuum temperature profile is $T_0(r)\propto r^{-0.75}$ 
        in agreement with the thin disk blackbody model, and, on top of it, is a 
       parametrisation of the hotter AEI spiral with $\gamma$ representing the strength of the instability expressed as the temperature contrast between the spiral and the surrounding disk.
       As the spiral extends radially from its corotation radius, $r_c$, its strength fades following a power-law parametrized by $\beta$ 
       as is seen in all AEI simulations. The last term is the shape function $r_s$ of the spiral with its width being defined by $\delta$ times the corotation radius.
The shape function for a spiral is:
      \begin{equation}
      r_s(t,\varphi) = r_c \,\mathrm{exp}\left(\alpha (\varphi-\Omega(r_c)t)\right)
      \end{equation}
where $\alpha$ is the spiral opening angle and $\Omega(r_c)=\sqrt{GM}r_c^{-1.5}$ is the Keplerian\footnote{in the case of the AEI $r_c$ 
is always far enough from the last stable orbit for the rotation curve to be well approximated by the Keplerian case}
frequency at $r_c$. Ultimately it is the rotation frequency of the spiral and the frequency at which the flux is modulated.
\\

 Once we have the temperature profile of the entire disk we assume it emits as a blackbody at the temperature $T(t,r,\varphi)$ namely  following 
      \begin{equation}
      B(\nu,T) = \frac{2h^2\nu^3}{c^2}/(e^{\frac{h\nu}{k_bT}}-1)
            \end{equation}
             with $k_b$ the Boltzmann constant, $\nu$ the frequency, c the speed of light, h the planck constant. 
        Hence the specific intensity emitted at some  position in the disk is  $I_\nu^{\mathrm{em}} = B_\nu(\nu^{\mathrm{em}},T)$  with ${em}$
     refering to the emitter's frame.
     Which then needs to be computed in the distant  observer's frame $I_\nu^{\mathrm{obs}} = g^3 I_\nu^{\mathrm{em}}$ 
     with the redshift factor  $g=\nu^{\mathrm{obs}}/\nu^{\mathrm{em}}$. 
       This is done with the open-source general relativistic ray-tracing code  GYOTO~\citep{Vin11}
  	into which we added our disk profile  defined by Eq.\ref{eq:T}. 
	
 

   \subsection{Timing studies and the shape of the pulse profile}            
    
       	A direct benefit of the mass scaling is the longer timescale for the modulation going from sub seconds in microquasars to hours and even hundreds of 
	days for the most extreme mass black-holes. And while such long timescales mean we cannot  directly apply the same technics as in LMXB, it also means we have access 
	to 
	\refe{new}  observables \refe{to which the existing models have not been tested against, hence giving us new ways to compare theoretical predictions and observations.}  

\begin{figure}[t]
\centerline{\includegraphics[width=90mm]{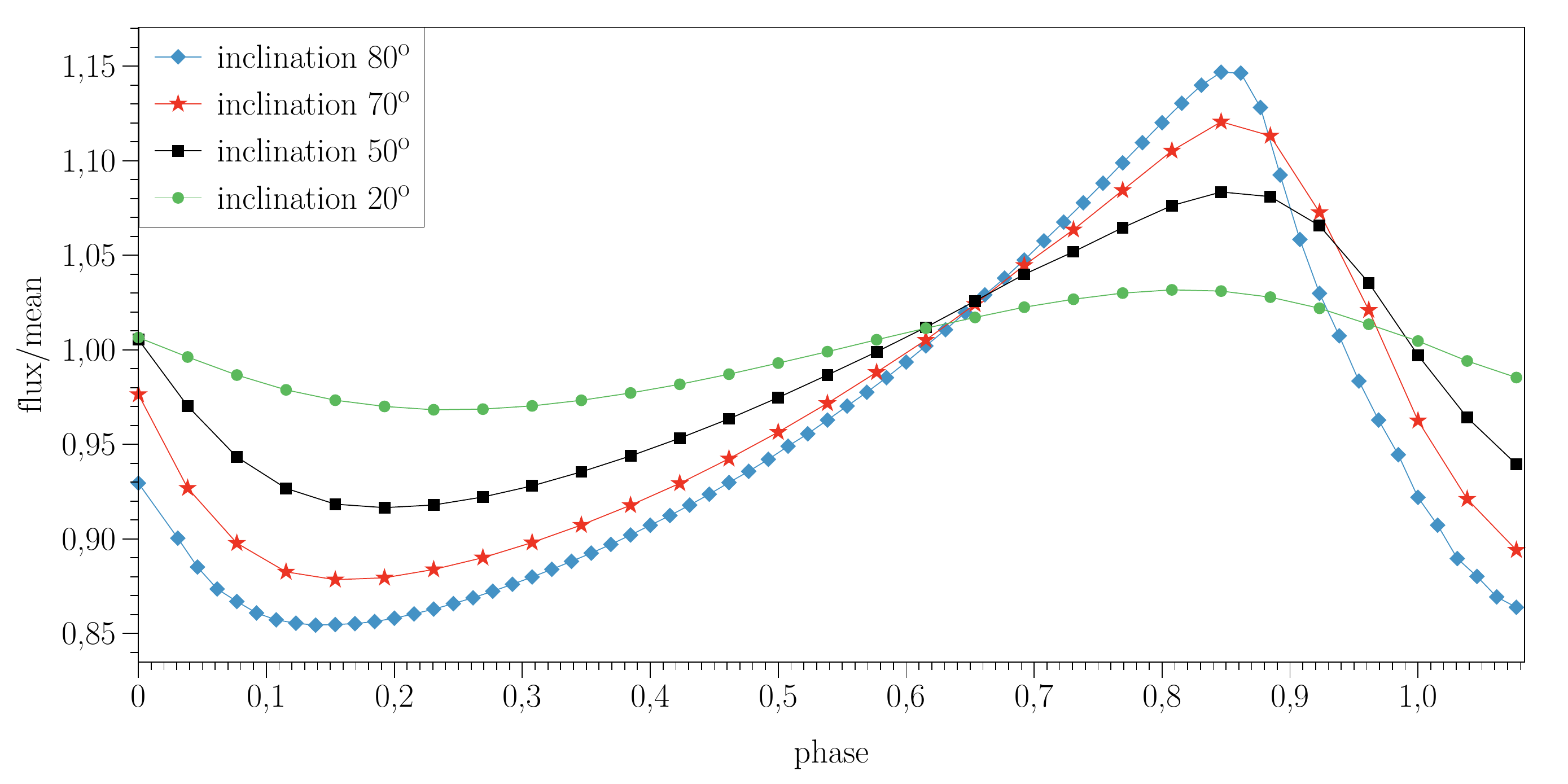} }
\centerline{\includegraphics[width=90mm]{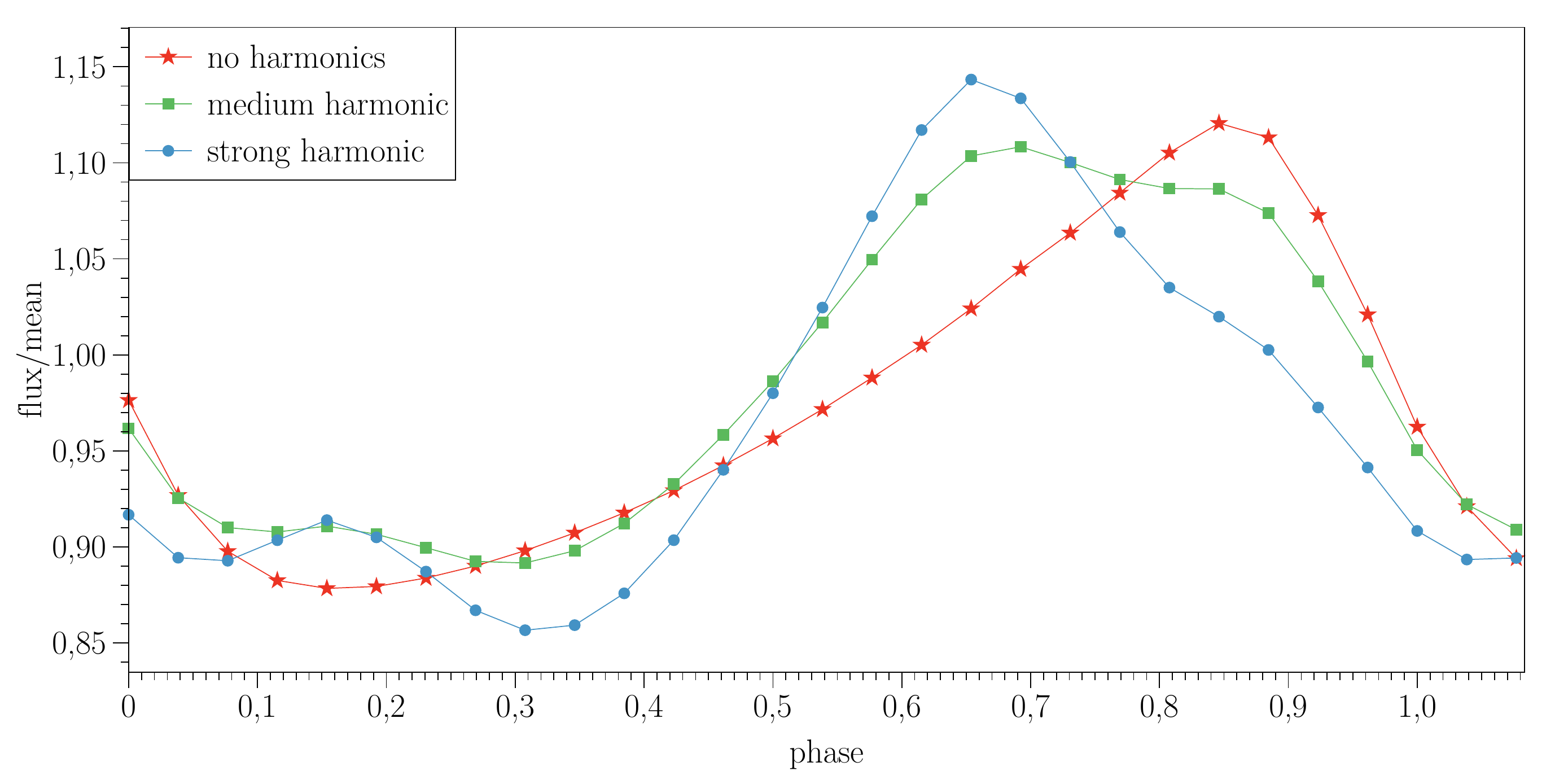}  }
\caption{Upper Fig: comparison between the pulse profile created by a  spiral in rotation for a disk around a Schwarzschild black hole seen at different inclinations. 
Lower Fig: impact of the presence of a strong or medium harmonics on the pulse profile created by a  spiral in rotation for a disk around a Schwarzschild black hole seen at 70$^\circ$.
For more details look at  sec.\ref{sec:toymodel} and \cite{VV16}.\label{fig:spiral_pulseprofile}}
\end{figure}

      Indeed, the hard to obtain QPO pulse profiles in the case of LMXB\citep{morgan97,2005A&A...434L...5V} becomes much more easily attainable when we go to the longer timescale of SMBH. 
        As an expression of the real-time modulation of the flux, it gives us access to 
        \refe{new} information about not only the origin of the modulation but also its geometry and can be used to 
        distinguish between models.    \\
       Thanks to the model presented in the previous section,  we can produce the pulse profile that would come from having the AEI in the disk and see how it is influenced by different parameters of the systems.  
        For example, in the case of the AEI, the shape and amplitude of the pulse profile have a strong dependance on  the inclination of the system as shown on the top plot of Fig.\ref{fig:spiral_pulseprofile}.
        While SMBH systems do not always have a well constrained inclination, by fitting the pulse profile with the AEI we can get an independent measure of the acceptable  range  of inclination. 
        Or, in the case of a well constrained  system, we can use the shape of the pulse profile to differentiate between QPO models as they tend to predict different shapes of pulse profile at a given inclination\citep{VV16}. \\
        
     Another benefit of having a good resolution on the pulse profile is that we can directly see the presence of harmonic structures on its shape. Indeed, this is illustrated on the lower plot of Fig.\ref{fig:spiral_pulseprofile}
        with the blue circle representing the extreme case of a \lq cathedral\rq\  QPO where the fundamental and the harmonics have a similar amplitude, whereas the green square represents  a much more common case 
        where the harmonic has about half the rms amplitude of the fundamental.  While those two cases are clearly distinct from the case of the fundamental only, represented by the red star, it still requires us
       to obtain small enough error bars on the pulse profile to be able to differentiate them observationally. 
       This could be attained by folding the lightcurve at the frequency of the QPO which will be easier for SMBH as 1) we see the QPO directly on the lightcurve and 
       2) an observation is rarely more than a few pulses and therefore the frequency is not expected  to vary  during an observation as is often the problem in LMXB.      \\
       
       Lastly, if we have enough counts in several energy bins, we can also look at how the pulse profile changes in different energy bands. This is especially interesting as it will increase even further 
       the constraints on the mechanism at the origin of the modulation and gives us a new way to differentiate between models.

\subsection{Imprint on the energy spectrum}

	The main feature of the AEI is the spiral structure which is at the origin of the modulation of the flux. Interestingly this spiral is also denser and hotter than the underlying disk, 
	therefore its emission will cause a phase-varying departure from the standard multicolor disk. This means that the AEI has an intrinsic impact on the energy spectrum. 
	Up to now, this was not studied in details as there was no hope to detect directly such fast changes for LMXBs.
       Indeed, such direct impact of what causes the modulation on the energy spectrum is only hinted at in the case of microquasars. 
      Nevertheless, thanks to those indirect detections,  it was surmised in \cite{VMR16} that QPOs could imprint the X-ray spectra of black hole binaries.
      By going to much higher mass systems, the longer timescale is going in our favor and it could become feasible to follow the evolution of the energy spectrum
       along one oscillation of the QPO.
       
       Observationally this requires to get enough photons in several phase bins of the pulse profile to be able to not only produce an energy spectrum in several phase bins, but to also have small error bars to 
       be able to determine if a shift between the energy spectrums is present. It is therefore important to first assess quantitatively the impact of the AEI on the energy spectrum
       to see if this could be detectable if actively looked for. \\

\begin{figure}[t]
\centerline{\includegraphics[width=90mm]{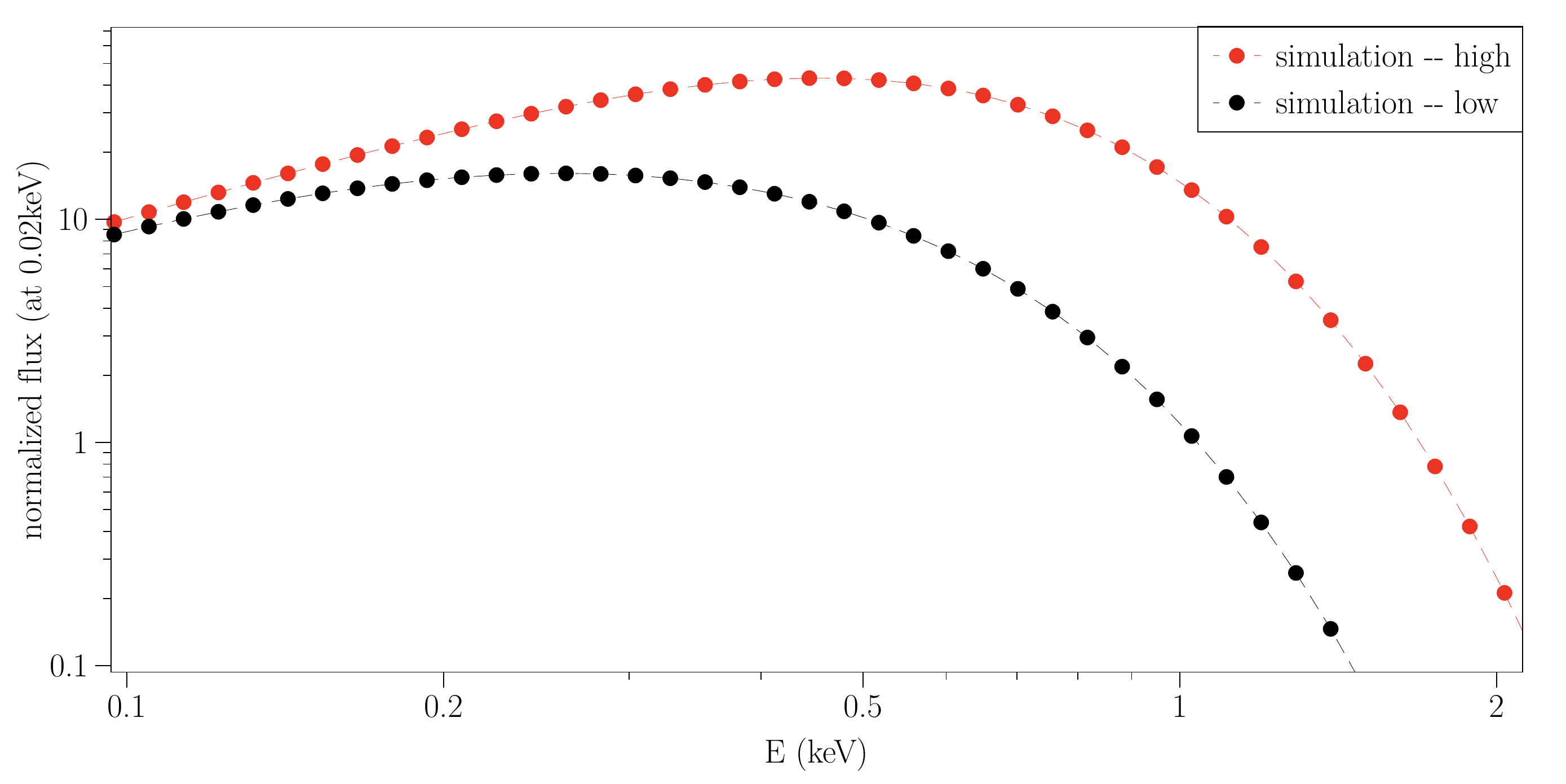}  }
\caption{Disk spectra comparison between the high(black circle) and low(red square) flux state of the disk in the presence of the AEI. To ease the comparison and show the shift, all fluxes are normalised to their common value at $0.02$ keV. The parameters are the one for 2XMM J123103.2 used in sec.\ref{sec:proof}. \label{fig:a0.9_diskspect}}
\end{figure}

                 Following the creation of the pulse profile in the previous section we are also  able to produce synthetic spectra for the  different phase bins along it, hence looking at 
                 how different the energy spectrum looks like depending on the phase of the spiral. 
      Fig.\ref{fig:a0.9_diskspect} shows the direct impact of the spiral wave in the disk on the energy spectrum of the full disk depending if we are at the low or high point in the pulse profile
       in the case of 2XMM J123103.2 (see sec.\ref{sec:proof}). To emphasize the difference both spectra are normalized by their common value at 0.02 keV.
       The shift shown between spectra here is an upper limit as using numerical simulations allows us to have at the same time a high number of photons and a very small phase bin. In the case of 
       a real observations the bin in phase will need to be larger to obtain enoughs photon to get good spectra and therefore those resulting spectra will not be as different. \\

       Nevertheless, thanks to the longer timescale of SMBHs the impact of the AEI on the energy spectrum can finally be put to the test. 
       By performing spectro-timing analysis where one model fits {\bf simultaneously} the pulse profile and how the energy spectrum changes along it, we can put more constraints on models and 
        drive home that QPOs are more than just timing features.

\section{proof of concept: spectro-timing analysis of 2XMM J123103.2}
\label{sec:proof}
     In order to do that, we turn to \refe{AGN} 2XMM J123103.2  (J1231+1106 after) \refe{a Narrow-line Seyfert 1} which exhibits QPOs similar to LFQPOs and for which both the 
     pulse profile and the energy spectrum associated with different parts of the pulse profile have been  published \citep{Lin2013}.      
      
             The aim of this proof of concept is three fold:
\begin{description}
    \item[\tt -]  First and foremost we aim to show in observation  the predicted behavior we presented earlier, where the energy spectrum changes along the pulse profile of the QPOs.

    \item[\tt -] Secondly, we aim to show that the AEI can indeed give a good representation of the pulse profile and the simultaneous changes of the spectral energy distribution,

    \item[\tt -] Lastly, the overall aim is to show the interest of producing both the pulse profile and how the energy spectrum behaves along it so that there are more objects with which 
    we can test QPO models.
 \end{description}

\subsection{3.8 hr Periodicity from an Ultrasoft AGN}    

     J1231+1106  was detected at off-axis angle in three XMM-Newton observations between 2003 and 2005.  The emission is detectable in the energy range of 0.2-2 keV.\\
  \begin{figure*}
\centerline{\includegraphics[width=95mm]{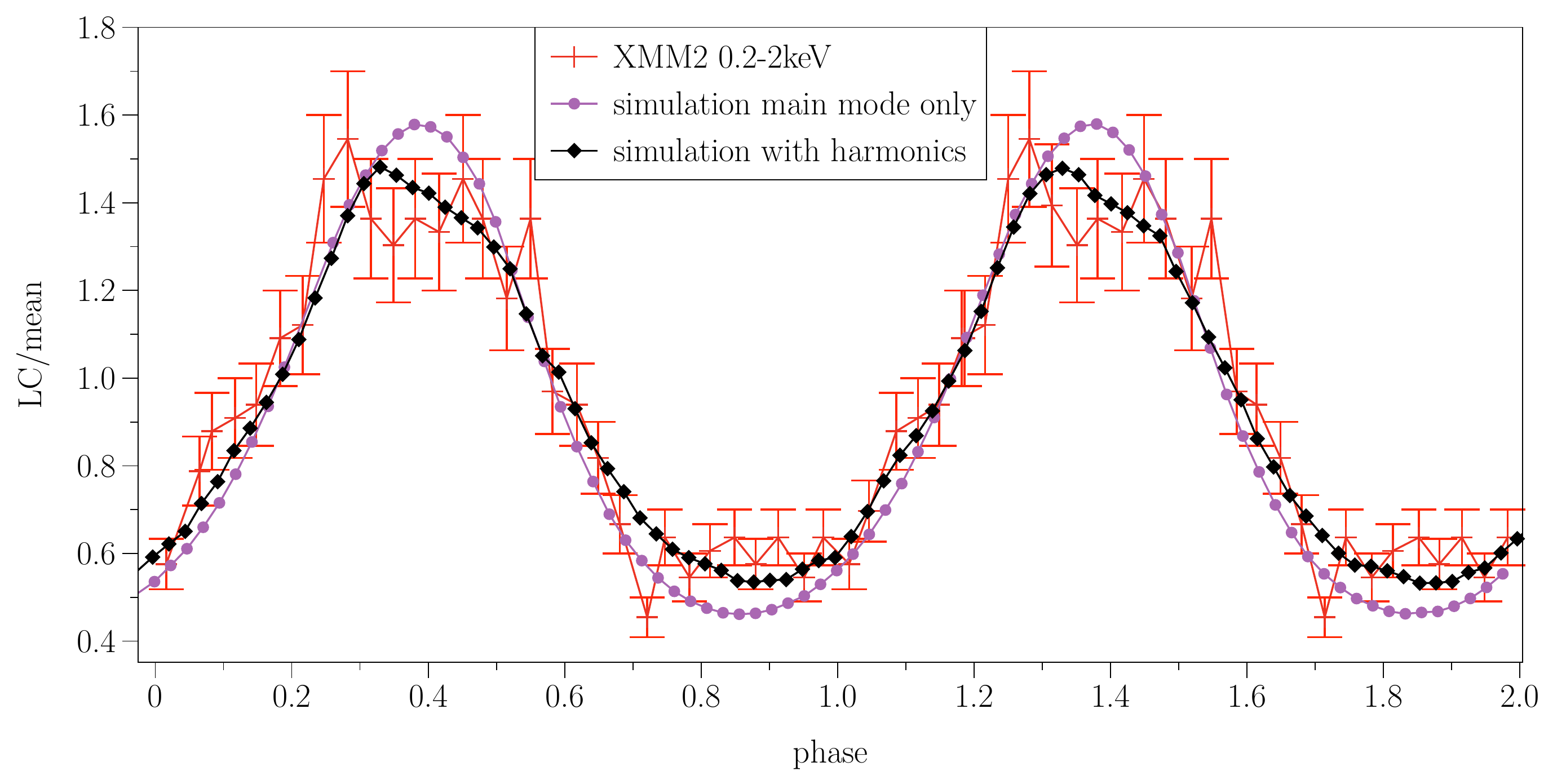}
\includegraphics[width=95mm]{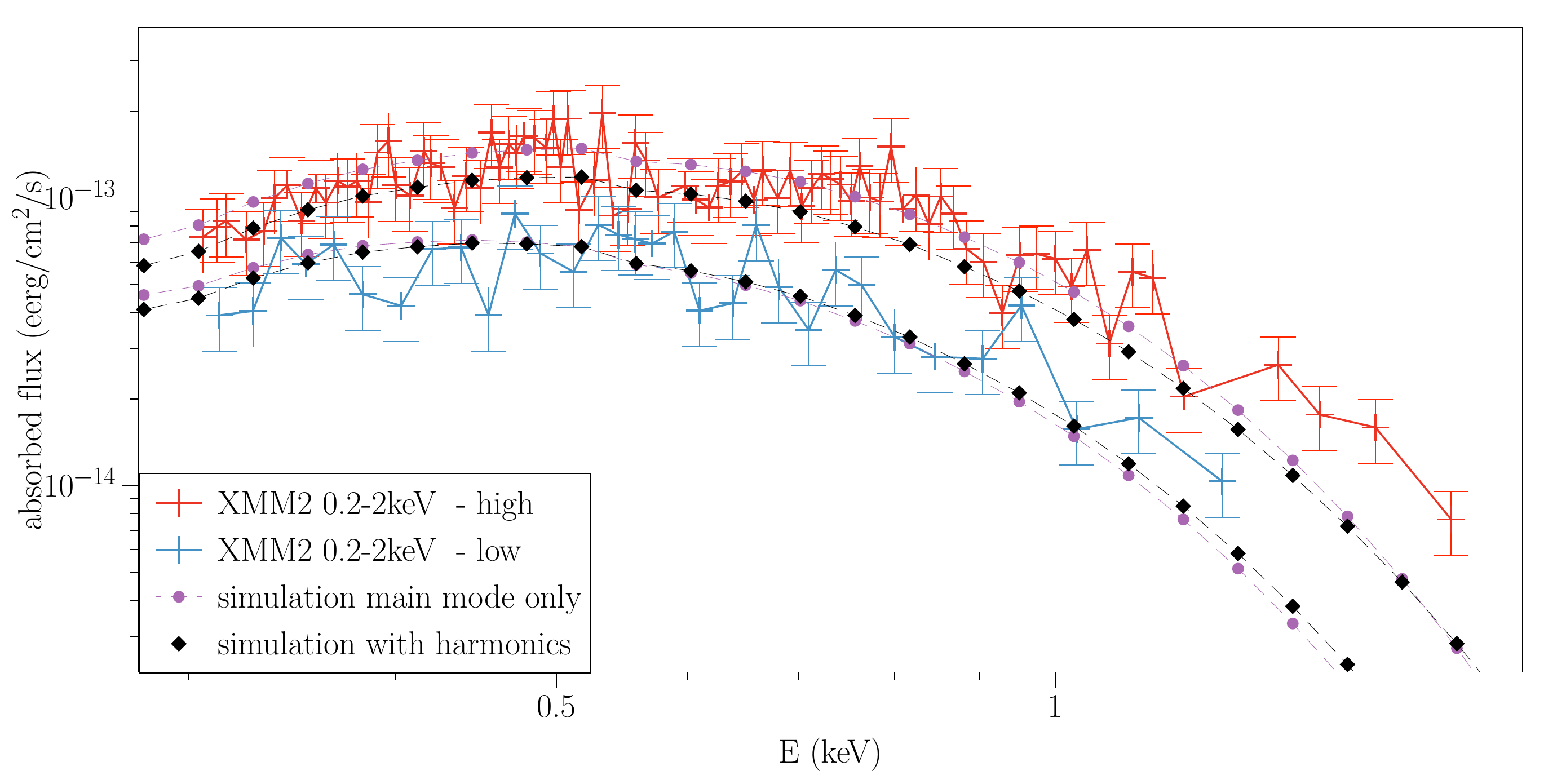} }
\caption{Left: Folded light curve from the XMM2 observation (red crosses) with the best fit from the AEI using the main mode of the oscillations (purple circle). A second fit,  
adding the harmonics that the shape of the pulse profile hints at is also given (black diamond).
Right: Energy spectrum  the XMM2 observation (red and blue crosses) with the best fit from the AEI using the main mode of the oscillations (purple circle). The energy spectrum associated with the 
the harmonics are also given (black diamond). \label{fig:fullXMM2_fit}}
\end{figure*}

     \cite{Lin2013} discovered a  large $\sim 3.8$h quasi periodicity in the two 
     XMM observations from december 2005  (XMM2 and XMM3) but no clear signal in the 2003 observation. 
     Those two observed QPOs have a significance of respectively $3.1\sigma$ and $4\sigma$ when assuming a power-law model for the red noise, and a very high rms 
     that increases from $\sim 25$\% in the 0.2-0.5keV to about $\sim 50$\% in the 1-2keV\footnote{When adding a QPO in the XMM1 observation it is consistent with a 0 rms,
     renforcing the idea that, even if present, the QPO is not detectable.}. \\

     Using the XMM2 observation, \cite{Lin2013} also looked at how the energy spectrum of the source changes during those oscillations. In that respect, they compared the 
     spectra found when taking only the \lq high flux\rq\  and  \lq low flux\rq\  intervals (with the cut-off being 0.05 count/s in the pn 1ks 0.2-2keV). While the spectra appear shifted in
     flux and the inferred fits are different, it is only at a $1.7\sigma$. 
     
      Because of the low detection level of the difference  it is not a firm confirmation of the predicted behavior by the AEI, nevertheless it is fully compatible with it and 
     a good raison to 1) try the spectral-timing fitting and 2) explore and search for more occurrences of SMBH QPOs with an even better statistic.

\subsection{Reproducing the observed behavior}

   The data used here (red crosses) come from the XMM2 observation in \cite{Lin2013} and in particular we focus on the pulse profile and the two 
    associated energy spectra for the high and low flux part of the lightcurve (Fig.\ref{fig:fullXMM2_fit}). 
     Here we aim to show that, with our toy-model presented in section \ref{sec:toymodel}, we are able to reproduce {\bf simultaneously} the 
     pulse profile as well as the energy 
     spectrum taken in the lower and upper sections of the profile. This is only a proof-of-principle as one would need not only more than one observation but also better statistics
      \refe{on the separation between the two spectra produced from that observation 
      \citep[which is at  $1.7\sigma $][]{Lin2013}, in order to fully confirm this behavior.}

     \refe{Because of those limitations and to not add too many unknowns to the fit, we explored first the simplest $m=1$ mode of the AEI, meaning the case of a QPO without harmonics.
     The results of those fits are shown as purple circles in Fig.\ref{fig:fullXMM2_fit}, and while the smaller substructure of the pulse profile, probably coming from an harmonic as shown in  Fig.\ref{fig:spiral_pulseprofile},
     it is able to reproduce the main observed features, such as the amplitude and phase of the pulse profile and the slight shift in energy spectrum associated with it.} 
  

        \refe{While no harmonic was detected in \cite{Lin2013} there are some  \lq below detection level\rq\  peaks in their PDS that could explain the }  substructure/departure from a  purely sinusoidal signal \refe{seen
        in the pulse profile. Hence,}
      we decided to look at what would the emission look like if this harmonic was present in the system at the level needed to reproduce the observed PDS.
    We see that the black diamond curve follows more closely the \refe{minute features of the} pulse profile while still able to explain the differences in the energy spectrum.    
    \refe{Hence, further investigation is needed to obtain better statistics on the observationnal pulse profile and how the energy spectrum evolves along it.}

\section{Conclusion}
 
     With the increasing number of QPOs being observed in super massive black-holes 
    we are looking at extending the LFQPO model based on the AEI to higher masses. While this is theoretically straightforward as the mass of the central object is only involved in the 
    scaling of the disk and the associated frequency, the difficulty here lies in the timescale change and what it implies for observations. \\
    
    Nevertheless, such long timescales can be used to our advantage as they lead to new sets of tests and constraints for models. Indeed, the AEI predicts that the energy spectrum will 
    change along the pulse profile and such change is potentially detectable if the rms of the QPO is strong enough. This led us to propose simultaneous spectro-timing fiting 
    of the pulse profile and the associated energy spectrum along it as a new way to test QPO models in super massive black-hole systems. 
    
    We presented the case of  J1231+1106  \citep{Lin2013}  as a potential detection of such change in the energy spectrum along the pulse profile of the QPO and also 
    demonstrated the ability of the AEI to simultaneously reproduce such behavior. This  last point is  just a proof of concept as there is only one \lq detection\rq\  at $1.7\sigma $.
     In order to fully test the prediction of the AEI there need to be more studies of the existing QPOs to produce both the pulse profile and how the energy spectrum behaves along it.\\

     In the long term, having ample cases of such study could help differentiate between QPO models. Indeed, up to now QPO models have focused mostly on explaining timing feature, 
     while ignoring their potential impact on the energy spectrum. Hence, having access to SMBH's longer timescale, we will be entering an era where QPOs are more than just 
     timing features in the PDS. 
    

\section*{acknowledgements}
Computing was partly done using the DANTE cluster in Paris at at the CINES in France. 
\bibliography{../../../../BiblioFull}
\end{document}